\documentclass[12pt]{article}
\usepackage{amsmath,amsfonts,amssymb}

\newcommand{\mathsym}[1]{{}}
\def\id{\protect{{1 \kern-.28em {\rm l}}}}

\def\be{\begin{eqnarray}}
\def\ee{\end{eqnarray}}
\def\p{{\partial}}

\makeatletter
\renewcommand\section{\@startsection {section}{1}{\z@}%
                                   {-3.5ex \@plus -1ex \@minus -.2ex}%
                                   {2.3ex \@plus.2ex}%
                                   {\normalfont\large\bfseries}}
\renewcommand\subsection{\@startsection{subsection}{2}{\z@}%
                                   {-3.25ex\@plus -1ex \@minus -.2ex}%
                                   {1.5ex \@plus .2ex}%
                                   {\normalfont\normalsize\bfseries}}

\makeatother

\def\Tr{{\rm Tr}}

\def \foot {\footnote}
\def \bi{\bibitem}
\def \tr {{\rm tr}}
\def \ha {{1 \over 2}}
\def \td {\tilde}
\def \ci{\cite}

\def \cN {{\cal N}}

\def\p{\phi}

\def\C{{\bf C}}

\def \del{\partial}

\def\g{\gamma}
\def\ov{\over}

\def \cL {{\mathcal L}}

\def\l{\lambda}

\def\foot{\footnote}
\def \four{{\textstyle {1\ov 4}}}
 
\def\det{\hbox{det}}
\def \ci {\cite}

\def \foot {\footnote}
\def \bi{\bibitem}
\def \ha {{1 \over 2}}

\def \fo { {1\ov 4}}
\def \ep {\epsilon}

\def \Tr {{\rm Tr}}

\def \l  {\lambda}

\def \td {\tilde}

\def \bi{\bibitem}
\def \la {\label}

\def \l {\lambda}
\def\foot{\footnote}

\def \sql {{\sqrt \l}}

\newcommand{\rf}[1]{(\ref{#1})}
\def \ov {\over}

\def\cc{\circ}

\def \ha{{1\ov 2}}

\def \no {\nonumber}

\def \del {\partial}

\def \cN {{\cal N}}

 \def \bb {\bar \beta}

\def \bi{\bibitem}
\def \la {\label}

\def \l {\lambda}
\def\foot{\footnote}

\def \sql {{\sqrt \l}}

\def \cL {{\cal L}}

 \def \p {\phi}

\def \ov {\over}

\def \varpi {{\rm w}}
\def \OO {{\cal O}}

\def \ep {\epsilon}

\def \te {\theta}

\def \cc {{\rm f}}

\def \cL {{\cal L}}

\def \C {{\cal C}}

\def\Tr{{\rm Tr}}

\def \del {\partial} 

\def \s {\sigma}

\def \G {\Gamma}

\def \p {\phi}

\newcommand{\dx}{\partial X}

\def \os  {\OO({\textstyle{ 1\ov \sql}})}

 \def \sql {\sqrt{\lambda}} 
\def \vp {\varphi}
 
\def \cc {{c }} 
\def \OO {{\cal O}}
\def \te {\textstyle}
\def \fl {\sqrt[4]{\l}}

\def \fo {{\textstyle{1 \ov4}}}
\def \rx {{\rm x}}
\def \hg {{\hat \g}}

\def \C  {{\rm C}}
\def \hC  {{\rm \hat  C}}
\def \dd  {{\rm d}}
\def \bb {{\rm b}}
\def \dDelta {2}
\def \sql {{\sqrt{\l}}}
\def \ed {\end{document}}
 \def \an {{\rm an}} \def \nan {{\rm nan}}

  \def \xx {{(0)}}

\def \xp {{x^+_s}}\def \xm {{x^-_s}} \def \tn {\tilde n}

\begin{document}


\overfullrule=0pt
\parskip=2pt
\parindent=12pt
\headheight=0in \headsep=0in \topmargin=0in \oddsidemargin=0in

\vspace{ -3cm}
\thispagestyle{empty}
\vspace{-1cm}

\rightline{Imperial-TP-AAT-2012-03}


\begin{center}
\vspace{1cm}
{\Large\bf  
On duality symmetry  in  perturbative quantum theory 

\vspace{1.2cm}

  }

\vspace{.2cm}
 { R. Roiban$^{d}$, 
 A.A. Tseytlin$^{e,}$\footnote{Also at Lebedev  Institute, Moscow. }
}

\vskip 0.6cm

{
\em 
\vskip 0.08cm
\vskip 0.08cm 
$^{d}$Department of Physics, The Pennsylvania  State University,\\
University Park, PA 16802 , USA\\
\vskip 0.08cm
\vskip 0.08cm 
$^{e}$Blackett Laboratory, Imperial College,
London SW7 2AZ, U.K.
 }
\vspace{.2cm}
\end{center}

\begin{abstract}
 Non-compact  symmetries of extended 4d supergravities  involve duality rotations of vectors 
and thus are not manifest off-shell
 invariances in standard ``second-order'' formulation. 
To study how such symmetries  are realised in 
the quantum theory we consider   examples in 2 dimensions   where vector-vector
 duality is replaced 
by scalar-scalar one. Using  ``doubled'' formulation, where  fields and their
 momenta  are treated on an 
equal footing  and  the duality becomes  a manifest  symmetry of the action 
(at the expense  of the Lorentz symmetry),  we argue that the corresponding 
on-shell quantum effective action or S-matrix are  duality symmetric as well as 
 Lorentz invariant. 
The simplest case of discrete $Z_2$  duality  corresponds to a  symmetry of the
 S-matrix under 
flipping the sign of the negative-chirality scalars in 2 dimensions 
or  phase rotations of  chiral (definite-helicity) parts of vectors in 4 dimensions.
We also  briefly discuss  some 4d vector models  and  comment on implications of our 
analysis for 
extended supergravities.

\end{abstract}

\newpage
\setcounter{equation}{0} 
\setcounter{footnote}{0}
\setcounter{section}{0}
\renewcommand{\theequation}{1.\arabic{equation}}
 \setcounter{equation}{0}
\setcounter{equation}{0} \setcounter{footnote}{0}
\setcounter{section}{0}

\def \os {O(\textstyle{ {1 \ov (\sql)^2}} )}
\def \ost {O(\textstyle{ {1 \ov (\sql)^3}} )}
\def \cc {{c }} 
\def \OO {{\cal O}}
\def \te {\textstyle}
\def \fl {\sqrt[4]{\l}}

\def \ha {{{\textstyle{1 \ov 2}}}}
\def \fo {{\textstyle{1 \ov 4}}}
\def \rx {{\rm x}}
\def \hg {{\hat \g}}

\def \C  {{\rm C}}
\def \hC  {{\rm \hat  C}}
\def \dd  {{\rm d}}
\def \bb {{\rm b}}
\def \dDelta {2}
\def \sql {{\sqrt{\l}}}

 \def \an {{\rm an}} \def \nan {{\rm nan}}
 \def \nm {\tilde n_{11}}
 \def \tn {{\tilde n}}  \def \uni {{\rm inv}}
\def \ttn  {{\bar n}_{11}}\def \ep {\epsilon} 

\def \G {\Gamma} \def \es {$E_{7(7)}$ \ }
\def \dx {\dot x}  \def \tx {\tilde  x} 
\def \cc {\chi}
\def \fo {{\textstyle {1 \ov 4}}}

\def \rE {{\rm E}}   \def \rB {{\rm B}}   \def \rA {{\rm A}}   
\def \tA {{\tilde A}}\def \cL {{\cal L}}
\def \cL  {{\hat \Omega}}

\def \vp {\varphi}
\def \teta {\tilde \eta} 

\def \tG{{\tilde G}}


\section{Introduction \label{intro}} 
Recent discussions of potential  divergences in 4d supergravities  brought to light the question of
how the  non-compact symmetries 
 \ci{jul0,jul,marcus} of  classical  ${\cal N} \geq 4$ supergravity equations of motion (involving 
transformation of  scalars combined with  a 4d  duality rotation of  vectors)  extend to quantum level 
and  which constraints 
 on counterterms they impose  (see  \ci{ack,elv,bos,stelle,bei,rk,ren}  and refs. therein). 
 Duality  symmetries involving vectors are specific to 4d extended supergravities 
 and  have their origin  (which still remains to be fully clarified) 
  in the dimensional reduction  from higher dimensional   supergravities.
  
 A prototypical  example of a relevant scalar-vector subsector of  (${\cal N} \geq 4$)   supergravity theory  is provided 
by  the following (Minkowski-space) 
 action\foot{More general actions   contain    $G/H$ scalar part  and several  abelian vectors 
coupled to the scalars.}
 \be     S=- \ha \int d^4 x \Big[  
   (\del_m \p)^2   +   e^{4 \p}  (\del_m \cc)^2   
  + \ha  e^{-2\p}  F^2_{mn}  + \ha   \cc     F^*_{mn} F^{mn}  \Big]   \ ,  
 \la{1}
 \ee  
where $ F^*{}^{kl} \equiv \ha  \ep^{klmn}  F_{mn}$ and $k,l,m,n=0,1,2,3$. 
The scalar  part  here is  $SO(1,2)/SO(2)$ sigma model; its global   invariance   under $SL(2,R) \approx SO(1,2) $ 
is promoted to the  invariance of the full equations of motion (written in first-order form)  when combined 
with  vector-vector duality transformation.
The simplest example of such a transformation is at $\cc=0$, when $\p \to - \p, 
\  \  A_m \to \td A_m $ with  $ \td F_{mn}=    e^{-2\p}   F^*_{kl}$.

As this  symmetry is  not  of a standard type, i.e.  it is  not a manifest local 
symmetry\foot{Here the phrase "local symmetry" is used for a symmetry whose transformation 
rules of fundamental fields do not involve inverse derivatives.
The duality   exchanging  field strength   with its   dual 
does not act locally  on   vector potential.} 
of the action \rf{1}, one may wonder how it is  reflected  in the corresponding 
quantum effective action or S-matrix.
An early 
 application of  the expected   symmetry under  ``$F \to F^*$'' (i.e.  $F \to \cos \epsilon\, F + \sin \epsilon\, F^*$) vector-vector duality 
was to constrain   the structure of the leading one-loop divergent terms  in the 
on-shell effective  action or S-matrix   of the  Einstein-Maxwell theory  \ci{dese}
 and the ${\cal N}=1$   supergravity coupled to a vector multiplet 
 \ci{ver}  (where the duality rotation is a symmetry if combined
 with a chiral rotation of the fermion in the vector multiplet, $\delta F_{mn} = \epsilon  F^*_{mn}, \ 
 \delta \lambda = \epsilon \g_5 \lambda$).
It was  argued that 
the observed absence of all $F_{mn}$-dependent one-loop 
UV divergent   terms except for the $T_{mn} T^{mn}$ one 
(where $T_{mn}$ is  the  
vector stress tensor)   
   may be attributed to the fact   that  this  is the only 
  duality-invariant term at this order.
  
  As was pointed out in \ci{des}, the duality symmetry of Maxwell theory   is naturally 
  realised in the phase space formulation (at the  expense  of manifest Lorentz symmetry). 
  It was  noted  in  \ci{gai} that   the duality invariance of the stress tensor $T_{mn}$ and thus 
of the corresponding Hamiltonian \ci{des}  should   imply the 
invariance of the S-matrix  (but this point was not 
elaborated on).

The   duality 
     symmetry acting also   on scalars   turns out to be  anomalous on a curved background 
    due to fermion \ci{div}  and vector \ci{marcus,dolg,olg,bos} 
   contributions, i.e. already the bosonic  theory \rf{1} and also  ${\cal N}=4$ supergravity \ci{marcus}
   have  $U(1) \subset SU(1,1)$  anomalies. \foot{Free quantum 
    Maxwell theory in  curved   background  has  a duality   symmetry anomaly reflected in 
    $ \langle FF^*\rangle = { 1 \ov 48 \pi^2} RR^* $ \ci{dolg}. Once coupled also to scalars, it leads 
    to duality-noninvariant terms in the  effective action
    containing an $RR^*$ factor.}
   Such anomalies  do not   affect, however,  the invariance of the
    leading UV divergent terms  but produce finite non-invariant terms in the quantum effective action.

If one integrates out  the vector field, i.e. if one considers the  effective 
action $\G$ depending only on the scalars,
then it is expected to be $SL(2)$ invariant. Indeed, 
performing the  vector-vector $A_m \to \td A_m $  
duality in the path integral  (by adding, as usual,  the Lagrange multiplier term, etc.  \ci{ftd})  one finds the 
same partition function with $\td A_m $   coupled to $SL(2)$ transformed scalars, implying that integrating 
out the vector should give 
an invariant functional of the scalars.\foot{This is no 
longer  true  in general  on a  curved 4d background:
one may get an ``anomalous''  local curvature coupling    analogous to 
the dilaton shift under scalar-scalar  duality  in 2d case  (see  Appendix A).} 

A non-trivial question is what happens if one keeps   both the scalars and the vector
as arguments of the effective action  or external states  in  the S-matrix.\foot{Keeping  all the fields 
on equal footing  may be  natural in supersymmetric theories.} 
A natural expectation  is that this  duality symmetry (in its form as defined on the 
 classical equations of motion)
 should be  present in the quantum effective action  
evaluated  on the 
equations of motion or  in the on-shell  S-matrix.
The precise meaning  of the action of this symmetry on the S-matrix (beyond its purely-scalar part 
 \ci{ack,elv,bei}) 
 is, however,  not immediately clear  and is to be defined.  
A far less obvious possibility (discussed in \ci{ren,rk,bn,car})  is that the quantum effective  equations derived 
from an  off-shell effective action should be covariant under a deformed  version of this duality.\foot{
This  condition is the same as the form-invariance (``self-duality'') 
of the effective action under the ``Legendre'' transformation from 
the original to the dual variables. 
Due to $U(1)$ gauge invariance  the effective action should depend only on the field strength 
and thus one can formally apply the duality transformation  but, in contrast to the case of the classical theory, 
 it is  not clear a priori  why 
this should lead  to ``self-duality'' of the effective action, i.e. to 
a duality-covariant  set of equations of motion.
For  a  discussion of  classical duality covariant   non-linear (supersymmetric) theories 
see \ci{kuz} and references there.}

A motivation behind the present paper is to try to clarify these  questions. 
The vector-vector  duality in 4 dimensions,  or more generally 
the $p$-form -- $p$-form duality  in $d=2p+2$  dimensions,    naturally acts on 
the phase space: the corresponding first-order action is duality-invariant \ci{des}.
Replacing momenta by spatial derivative of a  new (dual)  field,  one can rewrite 
the phase-space action as an action for a ``doubled'' set of fields in which the 
duality 
acts locally (without inverse spatial derivatives)  and 
is  a manifest off-shell symmetry.
 This is  achieved   at the expense of  Lorentz invariance; 
in its standard form, the Lorentz invariance is recovered on the equations of motion.
 Such a manifestly duality invariant  action was first written down 
in 2 dimensions  
 (following earlier work on chiral scalars \ci{fj} generalized to chiral p-forms in \ci{ht})
in  \ci{t1}.\foot{In string theory context the scalar-scalar   duality has
a  target-space  interpretation as  T-duality acting on  coupling functions. 
The $O(n,n)$ duality-symmetric form of sigma model equations of motion 
based on doubling of coordinates  was considered in \ci{duf}.}
This   action  for the ``doubled''   set of fields is 
describing  the same number of degrees of freedom 
as the original action (and an equivalent quantum theory) 
  but  is more suitable for addressing the
above  questions  about realization of duality at the quantum level. 
The corresponding ``doubled''    action  for vectors in 4 dimensions was constructed in \ci{ss}
and was generalized  to $\cN=8$ supergravity to obtain a manifestly 
 $E_{7(7)}$ invariant  action  in \ci{hil}. 
Detailed study of symmetry aspects  of quantum theory based 
on this ``doubled'' version of $\cN=8$ supergravity
(in particular, the absence of $SU(8)$ anomaly) 
 appeared in \ci{bos}  though the
 crucial question of preservation of 4d  Lorentz symmetry  at the quantum level  was not addressed.

As the  general issues with quantum realization of a symmetry involving  the duality 
are the same in any number  $d=2p+2$  of dimensions  here we shall concentrate on a  technically 
simpler but  still  non-trivial case  $d=2, \ p=0$,  i.e.   the case  when the 4d  vectors 
 are replaced by scalars.  The 2d analog of the action \rf{1} is 
\be     S=- \ha \int d^2 \s  \Big[  
   (\del_a \p)^2   +   e^{4 \p}  (\del_a \cc)^2   
  +   e^{-2\p}  (\del_a x_s)^2  
 +     \ep^{ab} \ep^{rs} \cc  \del_a x_r \del_b x_s   \Big]   \ ,  
 \la{12}
 \ee  
where $a,b=0,1$ and $r,s=1,2$. 
Like \rf{1}, this model  has the $SL(2)$  symmetry of the $ (\p,\cc$)
sector extended to the full set  of equations of motion provided 
it is combined with 2d duality transformation on the scalars $x_s$. 
We need at least $n=2$  scalars $x_s$  to have 
the $O(n,n)$ duality group  (acting on $x_s$ and their  ``momenta'')  big enough 
to contain the $SL(2)$ acting on  $ (\p,\cc$).\foot{Note that this  sigma model 
has constant  3-form strength $H_{\cc r s} =  \ep_{rs}$ and 
its metric  has  only one non-zero component of the Ricci tensor:
$R_{\p\p}=- { 3 \ov 2}$.  This model  is not conformal (the beta-function
 cannot be cancelled  by a linear in $\p$ dilaton background).
}

Similar sigma models with an ``external''  $y_i=(\p,\cc$)   and ``internal''  $x_s$
sectors were considered in the string theory context in \ci{cgf}. 
Integrating out $x_s$ one finds the $SL(2)$ invariant quantum theory for 
 ($\p,\cc$)  modulo local dilaton shift term \ci{bu,st}, but the realization of the 
duality  symmetry on the full set of fields at the quantum level is a priori non-trivial.

 We shall consider the 
 ``doubled'' formulation \ci{t1}  
in which the  duality in $x_s$ sector and thus the $SL(2)$  symmetry of 
\rf{12} is manifest but the  Lorentz symmetry is not  (but is  recovered 
on-shell).\foot{One may try to regularize the theory in 
such a way that the Noether current of the modified off-shell Lorentz 
 symmetry \ci{tw,t1} (reducing on-shell to the standard one) 
 is  conserved in the presence of quantum corrections.
Assuming such a regularization exists (i.e.
 if the modified Lorentz symmetry is non-anomalous) the on-shell
  quantum corrections should  be invariant under the 
 standard Lorentz transformations.}
 It turns out to be natural to split $x_s$ into its chiral parts 
 so that in the simplest case of $\cc=0$ 
the  duality  symmetry of the S-matrix  translates into a
symmetry under   flipping the sign of the  anti-chiral part
and the  sign of $\p$. 
Similar transformation   will apply to higher-dimensional models, e.g., in 
4d one would need  to flip  the sign  of the  anti-chiral part of  the vector field. 


Below we  shall  consider examples  of computations of  quantum effective actions for 
simple sigma models  with structure similar to  \rf{12}
or its truncations  and   demonstrating 
how the duality  is realised at the quantum level.  We shall concentrate on the most non-trivial discrete 
subgroup of the duality ($\p \to - \p$ for $\cc=0$) as generalization to continuous transformations 
does  not  bring new conceptual problems. 
Starting with the 
manifestly duality symmetric formulation we shall  check the presence of the 2d Lorentz 
invariance in the quantum on-shell effective action or in the S-matrix.

In section 2  we shall  consider an example  of a 2d model  analogous to \rf{12} 
--  the sigma model with euclidean AdS target space. We shall  describe its  duality-symmetric 
``doubled'' action   and discuss the action of duality symmetry on the  corresponding quantum effective action. 
We shall argue that the corresponding S-matrix is  both duality-symmetric and  Lorentz-invariant. 
In section 3 we shall  consider a different  example  of duality-symmetric theory  with a non-linear 
action  for a single 2d scalar that  can be written  in a standard   form  by introducing a coupling 
(similar to the $\phi$-coupling in \rf{12}) 
to a non-propagating auxiliary field. We shall compute the corresponding 1-loop effective action 
and demonstrate its duality symmetry. We shall also explain that  the quantum-corrected 
effective action remains duality-covariant if treated  in loop perturbation theory. 
In  section 4  we shall discuss the 4d action \rf{1}  and also the Born-Infeld (BI) action 
 and  briefly 
discuss implications of our analysis for $\cN=8$ supergravity.
In Appendix A we shall point out the possible  presence of  duality-non-invariant local terms 
when the theory is defined on a curved background. 
Appendix B  will  describe some details of the calculation of the 
1-loop effective action for the scalar theory of  section 3.

\renewcommand{\theequation}{2.\arabic{equation}}
 \setcounter{equation}{0}
\section{AdS sigma model: an example of 
duality-invariant theory  in 2 dimensions}

Let us consider the  following extension of  the $\cc=0$ truncation of 
the model \rf{12}:
a  sigma-model based on  euclidean $AdS_{n+1}$  metric 
($s=1,...,n$)
\be
\la{1.2}
ds^2 = d\p^2 + e^{-2\p}   dx_s dx_s  \ . \ee
2d duality for  all $x_s$ 
 maps this sigma-model  into itself  provided one also  does the 
coordinate transformation $\p \to -\p$ \ci{kts}.
 This transformation interchanges manifest (Noether) charges 
with an equivalent subset of  hidden charges (conserved due to the integrability of the model) \ci{rtw}; this  is  the origin of ``dual conformal symmetry'' \ci{bm,brtw}.

\subsection{Classical theory \label{ads_classical}}

Writing the sigma model action for \rf{1.2}   in ``first-order'' form  ($a=0,1$; \  $(\del_a x)^2 =- \dot x^2 + x'^2$) 
\be 
&& \ \ \ \ \  S(\p,x)= \ha \int  d^ 2\s \Big[ -(\del_a \p)^2 - e^{-2\p}  (\del_a x_s)^2  \Big]\la{1.3}\\
 &&   \to  \ \  S(\p,p,x) = \ha \int  d^2 \s \Big[ -(\del_a \p)^2 + 2 p_s \dx_s  -   e^{-2\p} x'^2_s-   e^{2\p}   p_s^2   \Big]
\la{1.4} \ee
and introducing a new field $\tx_s$  such that $p_s = \tx_s'$ we get 
the following duality-invariant action \ci{t1}
\be 
&& \hat S(\p,x,\tx)=\ha  \int d ^2\s \Big[ -(\del_a \p)^2 
+  \dx_s \tx'_s  + \dot {\tx}_s x'_s    -   e^{-2\p} x'^2_s   -   e^{2\p}   \tx'^2_s \Big]\la{1.5} \\
&& \ \ \ \ \ \  \ \ \ \ \ \ \ \ = - \ha  \int d ^2\s \Big[ (\del_a \p)^2 
-  \Omega_{IJ} \dot X^I X'^J     +   M_{IJ} X'^I X'^J   \Big] \ , 
\la{1.6}  \\
&& \ \ \ \    X=\begin{pmatrix} x \cr  \tilde x \end{pmatrix} \ , \ \  \  \ \ \ 
\Omega =    \begin{pmatrix} 0 & I\cr  I & 0 \end{pmatrix}\ , \ \ \ \ \ \ \ 
M =    \begin{pmatrix} e^{-2\p} & 0\cr  0 & e^{2\p} \end{pmatrix}\ , \la{1.7}
 \ee
where we used integration by parts  and $I,J=1,..., 2n$. 
This action  describes the same number of degrees of freedom as \rf{1.3} and is manifestly invariant 
under the duality transformation
\be  
\p \to -\p \ , \ \ \ \ \ \ \   x_s \to \tx_s\ , \ \ \ \ \   \ \ \ \tx_s \to x_s  \ . \la{1.8} 
\ee
Note that  the original sigma model action corresponding to \rf{1.2} is invariant under 
$SO(1,n+1)$, the dual action for $\p$, $\tx$ is also invariant under another  copy of  $SO(1,n+1)$ 
(which may be interpreted as part of  hidden symmetry of the original model). The interpolating 
``doubled''  action   \rf{1.5}  does not have  manifest $SO(1,n+1)\times SO(1,n+1)$ symmetry but 
it has of course an equivalent integrable structure (Lax pair)\foot{The corresponding flat currents 
in Poincar\'e-patch parametrization of $AdS_{n+1}$ were discussed in \ci{rtw}.
The Lax  pair corresponding to the 1st-order form of the equations of motion 
was  written down in \ci{brtw}. It is not, however,  manifestly  symmetric  under 
 \rf{1.8} as this duality should be accompanied  by a transformation of the spectral parameter 
 or $Z_2$ automorphism of the symmetry algebra \ci{brtw} 
 that implies certain transformation on the set of conformal charges.
 In particular, the Noether symmetry charges of the original model are mapped into hidden charges of the dual model and vice versa.}
 and thus the same set of conserved charges 
as the original $AdS_{n+1}$ model.\foot{The invariance under \rf{1.8}
corresponds to $ X\to \Omega X, \   M \to  \Omega M \Omega$.
This is  a special case $\Lambda =\Omega$ of more 
 general $O(n,n)$ transformations 
$X\to \Lambda X, \ \ \Lambda^T \Omega \Lambda = \Omega$,  that 
 preserve the structure of the action provided also $M \to  \Lambda^{-T}  M  \Lambda^{-1}$
 but this change of $M$  cannot be in general compensated by a redefinition of $\p$.
In the case of \rf{1.5},\rf{1.6} the continuous  part of  $O(1,1)$  duality 
symmetry can be realized by rotation of $X$ 
combined with translations of $\p$.}

Let us note that the analog of the  doubled action \rf{1.6} corresponding 
to \rf{12}   is given by 
\be 
&& \hat S(\p,x,\tx)=  -\ha  \int d ^2\s \Big[ (\del_a \p)^2 +
 e^{4 \p} (\del_a \cc)^2 
 -  \Omega_{IJ} \dot X^I X'^J     +   M_{IJ} X'^I X'^J   \Big] \ , 
\la{1.66}  \\
&&     X=\begin{pmatrix} x \cr  \tilde x \end{pmatrix} \ , \ \  \  \ \ \ 
\Omega =    \begin{pmatrix} 0 & I\cr  I & 0 \end{pmatrix}\ , \ \ \ \ \ \ \ 
M =    \begin{pmatrix} G- B G^{-1} B     &    B G^{-1}   \cr   - G^{-1} B     &           G^{-1} \end{pmatrix}\ , \la{1.77}\\
&& 
 (G- B G^{-1} B)_{rs} = (e^{-2 \p} +4 \cc^2 e^{2 \p} ) \delta_{rs} \ , \ \ \
( B G^{-1})_{rs} =2 \cc e^{2 \p} \ep_{rs} \ , \ \ \
G^{-1}_{rs}= e^{2 \p} \delta_{rs} \no   \la{88} 
 \ee
The symmetry of the full model  is  the $SO(1,2)$  subgroup  of $O(2,2)$ 
duality transformations on $M$ 
 that can be compensated by   $SL(2)$
transformations on $(\p,\cc)$.

The classical equations for $x_s$ and $\tx_s$  following from \rf{1.5} 
may be written as 
\be 
&&\ \ ( \dx_s - e^{2\p}   \tx'_s )'=0 \ , \ \ \ \ \ \ \ \ ( \dot \tx_s - e^{-2\p}   x'_s)'=0 \  \no \\
&&    \to \ \ 
\dx_s - e^{2\p}   \tx'_s =0 \ , \ \ \ \ \ \ \ \  \dot \tx_s - e^{-2\p}   x'_s =0 \ , \la{1.9} \ee
where as in \ci{t1} we  dropped  $\tau$-dependent integration functions assuming 
they are absent at the boundaries of spatial interval -- this ensures that we recover the standard equation of motion for $x_s$.\foot{Note that the action is invariant under 
$ \delta x_s = f_s(\tau).$
}

Introducing the combinations (that  become free  chiral  scalars  for $\p=0$)
\be  x_s = \xp + \xm \ , \ \ \ \ \ \   \tx_s = \xp -\xm \ , \ \ \ \ \ \ \ \ \ \ \ \ \ \ \ 
 x^\pm_s = \ha (x_s \pm  \tx_s) \ , 
 \la{1.10} \ee 
we can rewrite \rf{1.5} as ($\del_\pm = \pm \del_0 + \del_1$)  
\be 
&&\hat S(\p,x^+,x^-)= -\int d ^2\s \Big[ \ha (\del_a \p)^2  +    \xp' \del_- \xp + \xm' \del_+ \xm\cr
&&  \ \ \ \ \ \ \ \ \  \ \ \ \  \ \ \ \ \ \ \ \    \ \ \ \  \ \ \ \ \ \ \ \ \ \ \ \   +  f_1(\p) \ ( \xp'^2 + \xm'^2)  -  2 f_2(\p) \ \xp'  \xm'   \Big]\ ,  
\la{1.11}\\
&&\ \ \ \ \ \ \ \ \ \ \ \
f_1= 2 \sinh^2 \p \ , \ \ \ \ \ \ \ \ \ \  f_2 = \sinh 2\p\  \ .  
\la{1.12} \ee
The  corresponding  duality symmetry of this action  is 
\be \p\to -\p \ , \ \ \ \ \ \ \ \    \xp \to \xp \ , \ \ \ \ \ \ \ \ \   \xm \to - \xm    \ . \la{1.13} \ee
Like \rf{1.9} the  equations of motion for $x^\pm$  are Lorentz-invariant.\foot{The
``free'' action in \rf{1.5},\rf{1.11} is invariant under the  Lorentz-type symmetry:
$ \delta x_s = \tau x'_s + \sigma \tx'_s, \ \  \delta \tx_s = \tau \tx'_s + \sigma x'_s,$
or $\delta x^\pm_s= (\tau \pm \s) x'^\pm_s$. An analog of this  symmetry 
exists  also for non-zero $\p$ \ci{t1}. This symmetry becomes  standard Lorentz symmetry on 
the equations of motion.}

Note  that the  action \rf{1.11}  with any  even  $f_1$   and  odd $f_2$ functions 
would  be invariant under \rf{1.13}  but it will  be  Lorentz-invariant only for a 
special choice of $f_1,f_2$:
integrating $\tx$ out  one gets $  - (f_1 + f_2+1)^{-1} \dot x^2  + (f_1 - f_2+ 1) x'^2$ 
which is  Lorentz invariant only if $ (f_1 + f_2+1)^{-1} = f_1 - f_2+1 $.
Note also that $x^+_s$ and $x^-_s$  do not decouple in \rf{1.11}; they do if one considers 
$AdS_3$ and introduces a particular antisymmetric tensor coupling which leads to the $SL(2)$ WZW model.

\subsection{Quantum theory}

Let us now  turn to the 
quantum theory. The original  \rf{1.3} and the doubled theory \rf{1.5}
are expected to be equivalent quantum mechanically when probed with common observables.
An example of such observables are scattering amplitudes of $x$  fields, in which ${\tilde x}$ fields 
enter only through loops. Indeed, 
 integrating out ${\tilde x}$ in \rf{1.5} gives back \rf{1.3}.
The doubled theory allows, however,
 for a larger set of observables, e.g.  scattering amplitudes of both 
$x$ and ${\tilde x}$ fields 
which have duality acting as a  standard  symmetry.  
Given the  duality symmetry \rf{1.8},\rf{1.13} of the classical action \rf{1.5},\rf{1.11} 
one expects to find the same symmetry in the quantum effective action, i.e. 
\be
\G[\p,x,\tx]=  \G[-\p,\tx, x]  \ , \ \ \ \ \ \ \ \ {\rm i.e.} 
\ \ \ \ \ \  \ \ \G[\p,x^+,x^-]=  \G[-\p,x^+, -x^-]  \ . \la{g}
\ee
For this to  happen   one should  maintain the symmetry  at the quantum level by a 
proper choice of quantization prescription (i.e.  regularization and path integral measure). 
This may not be automatic if other fields and symmetries are also present.
 For example, the simplest special  case to consider 
would be  effective action depending just on $\p$ (found by integrating out both $x_s$ and $\tx_s$) 
which should  be invariant  
(by standard path integral  transformation argument \ci{bu,st})
  under just $\p\to -\p$. 
 As  we shall discuss in Appendix A, maintaining this duality 
 depends on assumptions about preservation of other symmetries (like target space
 diffeomorphism invariance), i.e.  on the  choice of measure and regularization scheme. 

The central  question,  however, is   if, like the classical action, 
the quantum effective action or S-matrix  will be  Lorentz-invariant  on-shell. 
This   on-shell invariance   may a  priori  have two    different interpretations:

(I)  $\G[\p,x,\tx]$  should be Lorentz-invariant once evaluated  on a solution 
of the  equations of motion, i.e., to the leading order,  \rf{1.9};  
 \foot{We will be  assuming for simplicity  that the relevant classical solutions 
 do not get non-trivial quantum corrections. 
 In a simple case of a field like $\tx$  entering the action only quadratically we may 
 expect that the corresponding equation of motion does not receive quantum corrections.
  In general, given a field theory with 
 a classical action  $S[\vp]$, the corresponding quantum S-matrix  generating functional
 is given by $\hat {\rm S} [\vp_{in}] = \Gamma[ \vp (\vp_{in})]$ where 
 $\Gamma[ \vp]$ is the quantum effective action  and $\vp (\vp_{in})$ is 
 the solution  of the {\it quantum} 
  equations of motion $ {\delta \G \ov \delta \vp} =0$ 
 with ``scattering''  boundary conditions, $\vp= \vp_{in} + ..., \ 
 (\del^2 + m^2 ) \vp_{in}=0   $ (see, e.g. \ci{jev}  and references therein). 
  While both should be invariant under the standard Lorentz transformations, 
  $\G$ evaluated  on the classical solution may differ from 
  $\G$  evaluated  on the solution of the effective 
  equation $ {\delta \G \ov \delta \vp} =0$ 
  starting with 2-loop order. 
%
 }

(II) the  quantum equations of motion following  from  $\G[\p,x,\tx]$   should be  
Lorentz-invariant.

\noindent
The property   (I)  should indeed be  expected given that the classical equations of motion 
are Lorentz-invariant   and that integrating over $\tx$ in \rf{1.5} leads us  back to the 
Lorentz-invariant action \rf{1.3}. This  may be verified explicitly at 1-loop order, i.e.
by expanding  \rf{1.5} to quadratic order in fluctuations  near a
classical solution $(\p_\xx,x_\xx,\tx_\xx)$  and restoring  the Lorentz invariance of the fluctuation Lagrangian by 
 a field redefinition of the fluctuation fields  which 
 makes the Lorentz invariance of the resulting effective action 
manifest.\foot{In view
  of the direct relation between \rf{1.4}   and \rf{1.5} 
  this is basically the same as the expectation 
 that in a semiclassical expansion with a phase-space action one should end 
 up with the same 1-loop effective action as found  in the usual  second-order Lagrangian formulation.}

The property   (II) is  likely to be true   too 
if  understood in a perturbative 
sense, i.e. that the effective equations of motion are solved  order by order in loop 
expansion: then given the classical  Lorentz invariance, the 1-loop corrected equations of motion should also be Lorentz-invariant, etc. 
However, (II) is far from obvious if  considered  as an exact property of the 
effective action: it is not a priori clear if  one should expect (some deformed version of)
Lorentz invariance to apply to the full quantum equations of motion. 
(II) is essentially equivalent to an  assumption  that quantum equations 
of motion derived from the original  Lorentz-covariant action \rf{1.3} 
should  admit an analog of the duality symmetry \rf{1.8}.\foot{The assumption that  such modified duality should apply to quantum counterterms 
in 4d supergravity was made in \ci{rk,ren}.  It is not clear, however, why this  property should apply  only to local (divergent)  part of the effective action. Also, if one is prepared 
to consider  subsets  of local  terms  in $\G$ forming duality-invariant   combinations (containing terms of all orders in fields/derivatives like BI action  and thus containing all powers of logarithm of a UV cutoff)  one may as well consider a possibility that UV divergent terms 
go away after a resummation of loop expansion.}

Since $\G[\p,x,\tx]$  evaluated on a general  classical solution with ``in''  (plane-wave)
initial  conditions at zero coupling is the generating functional for the S-matrix, 
 (I)  is equivalent to  the condition of Lorentz invariance of the 
the S-matrix for $\{\p,\xp,\xm\}$   following from \rf{1.11} 
(whose manifest duality invariance  \rf{1.13} is   expected 
 due to the   structure of the interaction terms with $f_1,f_2$ in  \rf{1.12}
but   2d Lorentz invariance is a priori non-trivial).
The key  fact is  that the on-shell conditions for the chiral scalars 
implied by \rf{1.11} are  Lorentz-invariant
\be  \del_- \xp =0 \ , \ \ \ \ \  \ \ \ \ \ \  \del_+ \xm =0 \ . \la{ccc}\ee
To demonstrate Lorentz invariance of the S-matrix  one  should note also that:

(i) to any loop order,  $x^\pm$  lines cannot ``terminate":
 they are either open (i.e.  $x^\pm$ coming into a diagram 
eventually exits it) or  closed (representing a loop of $x^\pm$ fields with $\p$ lines attached 
to it); 
(ii)  a tree-level Green's function with on-shell $x^\pm$ and off-shell $\phi $'s 
 is Lorentz invariant;\foot{To find  tree-level S-matrix one  may solve classical  equations of motion with ``in'' 
initial conditions, i.e. $x^\pm = x^\pm_{\rm in}  + ..., $ with 
 $\del_\mp x^\pm_{\rm in} =0$,  and substitute the result 
 into the classical action. This  gives a
  generating functional $\hat S(\p, x^+_{\rm in}, x^-_{\rm in})$
 for the corresponding tree S-matrix elements. 
Solving  the classical equation for the combination $\tx_s = \xp-\xm$ and substituting 
the result back into the action leads us to \rf{1.3}   with $x= \xp+ \xm$. The resulting functional
of $\xp,\xm$  is then  obviously Lorentz-invariant  (but of course is 
no longer manifestly duality-invariant).}
(iii)  the determinant of 
the $x^\pm$ -quadratic fluctuation operator depending on an off-shell $\phi$
  is Lorentz invariant.\foot{This follows again from the  fact that integrating out $\tx_s$ in \rf{1.5} leads us back to the  Lorentz-invariant  action \rf{1.3}. Note that this 
would  not be the case for  generic functions $f_1$, $f_2$ in \rf{1.11}.} 

\noindent
The observation (i)  allows us to break up any S-matrix element into parts  of  two  types 
that appear in  (ii)  and (iii)  which 
are connected by (Lorentz-invariant) $\phi$ propagators.  Since 
each part is Lorentz invariant,  the whole S-matrix element is  then also invariant.

 The same observations (i) and (ii) will apply of course in the  4d vector case of \rf{1} 
 where we
can split
 $F_{mn}$ into selfdual and anti-selfdual parts  that should correspond
 at the S-matrix level to positive and negative helicity photons; the (discrete part of) 
duality symmetry of the S-matrix will then mean a symmetry analogous to \rf{1.13}.

 We may also  compute the 1-loop S-matrix elements  explicitly  
and  check their  invariance under the duality transformations \rf{1.13} as well as their Lorentz
 invariance. 
 Albeit with a singular 
 momentum configuration, the simplest on-shell matrix elements are
 \be
 &&A(\phi(p_1),x_s^+(p_2),x_s^+(p_3))= A(\phi(p_1),x_s^-(p_2),x_s^-(p_3))=0 \ , \no 
\\
&& A(\phi(p_1),x_s^-(p_2),x_s^+(p_3)) \propto p_{2-}p_{3+}\,\ln\Lambda +\text{finite} \ ,\la{2.15}
 \ee
where $\Lambda$ is a UV cutoff. 
 As expected, they are Lorentz-invariant and renormalize the trilinear interaction in \rf{1.11}.\foot{We ignored the overall coupling constant in \rf{1.3}  which is logarithmically running in AdS sigma model.}
The matrix elements with four external $x_s^\pm$ are  the simplest ones with non-singular 
external momentum configurations.  It is easy  to see that the Feynman rules following from the action 
 \rf{1.11} do not allow four-point scattering amplitudes with an odd number of external $x_s^-$. For the 
 ones with an even number of external $x_s^-$ lines we find
 \be
&& A(x_s^+(p_1),x_s^+(p_2),x_s^+(p_3),x_s^+(p_4))
\cr
 &&  = \int \left[\frac{d^2 l}{(2\pi)^2}\right]_{\rm reg}
 \left[\frac{(p_1+l)_-}{(p_1+l)_+}+\frac{(p_2+l)_-}{(p_2+l)_+}\right] 
\left[\frac{(p_3-l)_-}{(p_3-l)_+}+\frac{(p_4-l)_-}{(p_4-l)_+}\right]
\frac{p_{1+}p_{2+}p_{3+}p_{4+}}{l^2(l+p_1+p_2)^2}\no
 \\[1pt]
 &&A(x_s^+(p_1),x_s^+(p_2),x_s^-(p_3),x_s^-(p_4))\cr
 &&= \ \int \left[\frac{d^2 l}{(2\pi)^2}\right]_{\rm reg}
 \left[\frac{(p_1+l)_-}{(p_1+l)_+}+\frac{(p_2+l)_-}{(p_2+l)_+}\right] 
\left[\frac{(p_3-l)_+}{(p_3-l)_-}+\frac{(p_4-l)_+}{(p_4-l)_-}\right]
\frac{p_{1+}p_{2+}p_{3-}p_{4-}}{l^2(l+p_1+p_2)^2}\no
 \\[1pt]
 &&A(x_s^-(p_1),x_s^-(p_2),x_s^-(p_3),x_s^-(p_4))\cr
 &&= \int \left[\frac{d^2 l}{(2\pi)^2}\right]_{\rm reg}
 \left[\frac{(p_1+l)_+}{(p_1+l)_-}+\frac{(p_2+l)_+}{(p_2+l)_-}\right] 
\left[\frac{(p_3-l)_+}{(p_3-l)_-}+\frac{(p_4-l)_+}{(p_4-l)_-}\right]
\frac{p_{1-}p_{2-}p_{3-}p_{4-}}{l^2(l+p_1+p_2)^2} \ .~~~~~~
\ee
As expected, these are also Lorentz-invariant provided that the regularization scheme included in the integration 
measure is chosen to preserve Lorentz symmetry.

 Let us mention  a potentially subtle issue of the choice of UV cutoff (or, more generally, the choice of 
 regularization scheme) in a theory without manifest Lorentz symmetry. While we do not expect a genuine 
 Lorentz anomaly in a theory with  balance of chiral spinors and self-dual tensors, there may nevertheless 
 exist a  spurious  breaking  of Lorentz symmetry due to an unfortunate choice of UV cutoff. 
 In general, one may  of course use an arbitrary  UV cutoff and  then attempt to add local 
 counterterms to satisfy the Ward identities of the required symmetries.  
 We expect the same philosophy  should   be applicable to the present case of on-shell Lorentz symmetry.
  

To try to check the possibility (II), i.e.  that the quantum effective action may 
have a nonlinear analog of the tree-level duality (cf. BI action  vs. Maxwell action)
one  may try to compute  the 1-loop  $\G=\G_1$  for a classically invariant theory
 like \rf{1} or \rf{1.3} 
in some approximation (e.g. keeping only field strength dependence but ignoring dependence 
on its derivatives). 
For example, starting  with \rf{1}  it is easy to find  $\G_1[F_\xx] $  for 
$F_\xx$=const. To have  a consistent  classical solution we will need to require that 
$\p=\p_\xx=$const  and thus $(F_\xx)_{mn} (F_\xx)^{mn}=0$.  In this  case $\G_1$ will 
depend only on the  traceless stress tensor or 
$T^k_n=e^{-2 \p_\xx} (F_\xx)_{mn} (F_\xx)^{mk}$ and thus   is (an even) 
 function of 
only one invariant $e^{-2 \p_\xx} (F_\xx)_{mn} (F^*_\xx)^{mn}$. It is   then   invariant 
under the classical duality symmetry  but  this approximation is not sufficient to address 
the question about possible  duality symmetry of the quantum equations following from $\G$: 
for that we need to know  the dependence of $\G$ on both $F_{mn} F^{mn}$ 
and $F_{mn} F^{*mn}$ invariants. 
A similar remark applies to the action \rf{1.3} where we may consider a classical solution 
with $\p=\p_\xx=$const, \ $\del_a x_\xx $=const  and $(\del_a x_\xx)^2=0$ (i.e. $\del_+ x_\xx=0$ or 
 $\del_- x_\xx=0$).  We shall discuss such a computation (and also its generalization)
on the example of a slightly  different scalar model in the next section.

\renewcommand{\theequation}{3.\arabic{equation}}
 \setcounter{equation}{0}
 \section{An example of   nonlinear scalar action}
 

Let us  now  consider a superficially different  but, in fact, related example of 
a   non-linear scalar  theory 
 depending only on $(\del x)^2$. It has 
   classical duality symmetry and we shall   study  if  this theory  has also 
   a generalization of the  duality at the quantum level. 
The  corresponding  action  is (here $x$ is a single scalar field and we ignore a dimensionfull coupling constant) 
\be   S=  \int d^2 \s \ L(x)  \ , \ \ \ \ \ \ \ \ \ \ 
 L(x)= -\sqrt{ 1 + (\del_a x)^2 }   \ . \la{hh} 
 \ee
Finding the  momentum $p$ conjugate to $x$  and setting $p\equiv  \tx'$
we get  the corresponding phase-space or ```doubled''  Lagrangian 
which is the   manifestly duality-invariant  analog of \rf{1.5} 
\be 
\hat L(x,\tx) =  \tx' \dot x -  \sqrt{ 1 + x'^2 } \sqrt{ 1 +  \tx'^2 } \ . \la{kk}\ee
Note  that here the integral over $\tx$   (or  the momentum) is non-gaussian 
so the  quantum theories   defined   by   \rf{hh} and \rf{kk} are equivalent 
only in the leading semiclassical approximation  of the integral over  $\tx$. 

Semiclassically,  \rf{hh}  is   equivalent to the following  Lagrangian
\be 
&& L(x,G) =  -  \ha \Big[ G (\del_a x)^2   +  G + G^{-1} \Big ] \ , \ \ \ \ \ \ \ \ \ \ \    G= e^{-2\p}
\ ,   \la{ll}\ee
where $G$ (or $\p$)  is an auxiliary  2d field.\foot{This  representation is a simple analog of replacing 
 the Nambu action with  the  ``Polyakov''  action with an independent  2d metric. 
 Analogous ``polynomial'' representations using auxiliary scalars exist for other  similar actions like Born-Infeld one  \ci{rot, rot1}.} 
The  corresponding ``doubled''   action is  then  the same as in  \rf{1.5}  
\be 
 \hat L(x,\tx) =  \tx' \dot x - \ha G (1 + x'^2  )  - \ha     G^{-1}   (1 + \tx'^2) \ .  \la{pl}\ee
The duality symmetry  of the equations of motion for  \rf{hh}
($x \to \tx $   with $\ep^{ab} \del_b \tx = [1+ (\del_a x)^2]^{-1/2}   \del^a x $) 
corresponds to 
$ x\to \tx, \  G\to G^{-1}$  which is the manifest symmetry of \rf{pl}.
Solving for $G$ in \rf{pl} leads again to \rf{kk}, while integrating out $\tx$ gives   back 
\rf{ll}. 

We thus get 
an analog  of  \rf{1.3}    but with a potential instead of a  kinetic term  for $\p$. 
The Lagrangian  that generalizes  both \rf{1.3} and \rf{ll} 
\be L= - \ha (\del_a \p)^2  - \ha e^{-2 \p} (\del_a x)^2  - \cosh 2 \p
\ ,   \la{ty} \ee
 also represent a duality-covariant theory.

In view of   non-polynomiality of \rf{hh}  it is natural (as in the Nambu $\to$ Polyakov  action case) 
to define  the corresponding 
 quantum theory  by the path integral  with the action \rf{ll} or the equivalent ``doubled'' action  \rf{pl}. 
If we start with \rf{ll}  and  integrate out $x$, we get  an effective  action for $\p$ 
 which is  invariant 
under $\p\to -\p$.  If we keep a background for $x$ and evaluate the effective action 
on shell, we should get again a duality-symmetric result as in \rf{g}. 
The classical solution  for  $x_\xx, G_\xx$ in \rf{ll}  satisfies 
\be 
    G_\xx n^a =  \ep^{ab} \tn_b  \ , \ \ \ \ \ \   n_a \equiv \del_a x_\xx  \ , \ \ \ \ \ \ \ 
 G_\xx = (1 + n^2)^{-1/2} = (1 + \tn^2)^{1/2} ={\td G}^{-1}_\xx  \ ,\la{clas}   \ee
where $\tn_a,\tG_\xx$ is the classical solution following from the action  dual to \rf{ll}
\be 
\td  L(\tx,G) =  -  \ha \Big[ G^{-1}  (\del_a \tx)^2   +  G + G^{-1} \Big ] \  . \la{dac} \ee
To compute the effective action we need  to expand   near the classical solution,  
$ \ x=x_\xx + \eta, \  G=G_\xx(1+ \xi)$.  If we  first treat $G$ as an external background and 
perform the path integral  duality transformation with respect to the  fluctuation $\eta$, we end 
up   with \rf{dac}  with $\tx = \tx_\xx + \teta$, where $\teta$ is dual to $\eta$. 
Integrating then over $\eta,\xi$ in \rf{ll}   or over $\teta,\xi$ in \rf{dac} we should get the same result for the effective action. 

Let us demonstrate this more explicitly. 
Expanding \rf{ll} 
we find the following quadratic-fluctuation action
\be 
L_2(\xi,\eta)= - \ha G_\xx (\del_a \eta)^2  - G_\xx  n^a \xi  \del_a \eta  - \ha G^{-1}_\xx \xi^2 \ .
\la{oo}\ee
Integrating out $\xi$ gives
\be 
L_2(\eta) = - \ha G_\xx (\del_a \eta)^2   + \ha  G_\xx^3  (n^a \del_a \eta )^2   \ .
\label{quads}
\ee
We may  now perform the standard  path integral  duality  over $\eta$   by  replacing  \rf{quads} with 
\be 
L(B,\teta) = - \ha G_\xx  B_a^2   + \ha  G_\xx^3  (n^a B_a)^2  
  +   \epsilon^{ab} B_a \del_b \teta\ ,
\label{qads}
\ee
Performing the Gaussian integral over  the auxiliary vector field $B_a$   we get  the  following  
 action for the dual  fluctuation field $\teta$ 
\be 
\td L_2(\teta) = - \ha G^{-1}_\xx (\del_a \teta)^2  
 + \ha  G_\xx^{-3}  (\tn^a \del_a \teta )^2   \ .
\label{quas}
\ee
This  is exactly the same quadratic-fluctuation action that  follows  
if one    starts with  \rf{dac}.
This   shows again  that the  resulting  effective action $\G_1$ 
  which is a functional   of  $ x_\xx, G_\xx$, i.e.
(in view of \rf{clas})  a 
 functional of  $n_a=\del_a x_\xx$  is duality-symmetric, i.e. invariant under $n_a \to  \tn_a$,
$G_\xx \to \td G_\xx= G_\xx^{-1}$.

This formal argument ignored local  measure-like factors
(like  a ``determinant''  term in $\G_1$ proportional to $\ln G^{-1}_\xx$ coming from integration over $\xi$)
which are absent in a regularization ignoring power divergences. 
In  general, the duality invariance  of the resulting  effective action depends 
on a choice of  measure/regularization. 
If we use a dimensionfull (e.g. proper-time) UV
 cutoff $\Lambda$
then the effective action 
 is  sensitive to the
contribution of the measure. The  standard  choice $ [dx] = dx \sqrt{G}$ 
(corresponding to the trivial measure 
in the  phase-space path integral  or path integral corresponding to the ``doubled'' action \rf{pl})
ensures the duality invariance of the result.\foot{We  also assume that 
the fundamental variable  is $\phi= - \ha \ln G $,    
i.e. the measure of integration over $\phi$ is trivial.} 

Let us consider    the  special case of $n_a=\del_a x_\xx=$const.
\foot{Below for notational simplicity we shall omit
subindex $_\xx$ on $x$ and $\tx$.}
Taking into  account the ``determinant'' term  from the $\xi$ integration 
and the measure contribution for $\eta$ the resulting 1-loop on-shell 
effective action may 
be written as 
\be \la{kkk}
\Gamma_1 = \ha \ln \det  K   \ , \ \ \ \ \ \ \ \ \ \ 
K = G^{-1}_\xx \del^a \del_a   -    G_\xx    (n^a \del_a)^2   \ . 
\ee
Using that $\ep^{ab} \tn_b   = G_\xx n^a $,
$ G^2_\xx = (1 + n^2)^{-1} = 1 + \tn^2 $  where  $ \tn_a=  \del_a \tx$, 
one can see  that 
\be K = G_\xx \del^a \del_a   -    G^{-1} _\xx    (\tn^a \del_a)^2   \ , \la{kky} \ee
 and thus $\G_1$  is duality invariant under $ x\to \tx, \  G\to G^{-1}$.
The resulting  (classical plus one-loop) effective action for constant $n_a = \partial_a x$ 
has the form (see Appendix B)
 \be 
 \G = \int d^2 \s \Big[ -\sqrt{ 1 + (\del_a x)^2 } 
 +   \Lambda^2  F\big(\sqrt{ 1 + (\del_a x)^2 }\ \big) \Big] \ , 
 \label{2.21}
 \ee
 where  the function $F(y)$ (whose argument  on-shell is    $G^{-1}_\xx$)
 is 
 \be  F(y) = \ln [ \ha ( y^{1/2} + y^{-1/2}) ] \ . 
    \la{ppp} \ee
 Its   symmetry   
 $F(y)= F(y^{-1})$  
 makes  the duality invariance of the  1-loop effective action  manifest.

Assuming one starts directly with \rf{hh}  let us now 
 study  whether the tree-level action plus the part of the 1-loop effective 
 action which depends only on the first derivative of $x$, 
\be
\Gamma(\del x)  &=&
-\int d^2\sigma\sqrt{1+(\partial_a x)^2}+
 \hbar\,\Gamma_1(\partial x)  +{\cal O}(\hbar^2) \ ,
 \ee
  leads to the  duality-covariant  quantum equations of motion
 to the relevant leading order in the loop expansion, i.e.
 is ``self-dual'' under the ``Legendre'' transform from $x$  to the dual variable.
From the  above discussion  the resulting $\Gamma_1$  is the same as found  
by starting from \rf{ll}   and it 
depends on $\partial _a x$ only through 
$G^{-1}_\xx=\sqrt{1+(\partial_a x)^2}$. 
To check the duality covariance 
 one is to carry out the  ``Legendre'' transform from the original  to the dual 
variable  while keeping all the relevant   ${\cal O}(\hbar)$ terms.
Replacing $\del_a x $ by an independent field strength 
 $n_a$   
 and introducing the dual variable $\tx$ through the 
 Lagrange multiplier term we get 
\def \xx {{(0)}} \def \yy {{(1)}} 
\be
\hat \Gamma (n,\del \tx) =
-\int d^2\sigma\sqrt{1+n_a^2}+
\hbar \Gamma_1(n)+{\cal O}(\hbar^2)
+ \int d^2\sigma\;\epsilon^{ab}n_a\partial_b{\tilde x}\ . 
\label{2.32}
\ee
Solving  the resulting effective   equation  for $n_a$ 
\be
\frac{n^a}{\sqrt{1+n^2}}- \hbar\frac{\delta\Gamma_1}{\delta n_a}+{\cal O}(\hbar^2) 
= \epsilon^{ab}\partial_b {\tilde x}
\label{2.33}
\ee
perturbatively in $\hbar$   we get 
\be
 && n^a=n^a_\xx+\hbar\, n^a_\yy+{\cal O}(\hbar^2) \ , \\ 
&&n^a_\xx = \frac{\;\epsilon^{ab}\partial_b 
{\tilde x}}{\sqrt{1+ (\partial_a {\tilde x})^2}} \ , \ 
\qquad
(1+n_\xx^2)^{-1/2} =\sqrt{1+(\partial_a {\tilde x})^2 }
\ , \\  
 &&
\frac{n^a_\yy}{(1+n_\xx^2)^{1/2}}  -\frac{n^a_\xx(n_\xx\cdot n_\yy)}{(1+n_\xx^2)^{3/2}}+ \Big(\frac{\delta\Gamma_1}{\delta n_a}\Big)_{n_\xx}=0 \ ,
\label{2.37}\\
&& 
n_\xx\cdot n_\yy = (1+n_\xx^2)^{3/2}n_\xx^a
 \Big(\frac{\delta\Gamma_1}{\delta n_a}\Big)_{n_\xx} \ , 
\quad
n_\yy^a = (1+n_\xx^2)^{1/2}(n_\xx^an_\xx^b-\eta^{ab})
 \Big(\frac{\delta\Gamma_1}{\delta n^b}\Big)_{n_\xx}
\no
\ee
The dual action, i.e.   \rf{2.32} expressed in terms of $\del \tx$   is then
\be
\td \Gamma(\del \tx) &=& -\int d^2\sigma\frac{1}{\sqrt{1+n_\xx^2}}- \hbar \int d^2\sigma\frac{ n_\xx\cdot n_\yy}{(1+n_\xx^2)^{3/2}}
+\hbar \Big(\Gamma_1-n_a\frac{\delta\Gamma_1}{\delta n_a}\Big)_{n_\xx} 
+{\cal O}(\hbar^2)
\cr
&=&-\int d^2\sigma \sqrt{1+(\partial_a{\tilde x})^2}+\hbar
\Gamma_1\Big|_{n_a\rightarrow \frac{\epsilon^{ab}\partial_b {\tilde x}}{\sqrt{1+(\partial_a {\tilde x})^2}}} 
+{\cal O}(\hbar^2)\ . 
\ee
As we have shown  above,
 the 1-loop effective action $\G_1$  is invariant under the 
classical duality transformation ($ \G_1 (\del x) = \G_1(\del \tx)$) 
 so that  the  $\td \Gamma(\del \tx) $  has the same form as 
 $ \Gamma(\del x) $, 
  up to ${\cal O}(\hbar^2)$ terms, 
\be
\td \Gamma(\del \tx)=-\int d^2\sigma \sqrt{1+(\partial_a{\tilde x})^2}+\hbar
\Gamma_1\Big|_{n_a\rightarrow \partial_a {\tilde x}} 
+{\cal O}(\hbar^2) \ . 
\label{3.23}
\ee
Since this argument  used only the duality-invariance of $\Gamma_1$,
 it follows,  quite generally, that if 
the leading quantum  correction to a classically ``self-dual'' (in the sense of the 
above ``Legendre'' transform)
 action is duality-invariant, then the resulting effective action is ``self-dual'' 
up to  higher-order corrections. 
%
%
The relation between the original and dual fields receives, in general,
 loop corrections  (cf.  \rf{2.37}).
%

One may attempt to  extend the above discussion 
 by including higher-loop corrections $\G_n$ to the 
  effective action and finding the constraints $\G_n$
must satisfy for the complete effective action to be
 ``self-dual'' through the required order.  For example, 
the two-loop effective action should be a solution of 
\be
\Gamma_2(\partial {\tilde x}) = 
\Gamma_2(n_\xx) + \int d^2 \sigma \frac{1}{(1+n_\xx^2)^{1/2}}\Big[n_\yy^2 - \frac{(n_\xx
\cdot n_\yy)^2}{1+n_\xx^2}\Big] \ .
\label{3.24}
\ee
It  is not clear a priori why $\G_2$  should 
obey this constraint.

\renewcommand{\theequation}{4.\arabic{equation}} \setcounter{equation}{0}
\section{Four-dimensional  vector models}

Duality symmetries (in 2d or 4d)  discussed above are on-shell symmetries; it is not possible to 
promote them to manifest  symmetries of the 
action while preserving all the other symmetries of the theory,
in particular Lorentz invariance. 
The ``doubled'' formalism provides a framework in which the duality symmetry becomes  a manifest
off-shell symmetry; while the  ``doubled''  action 
is not invariant under the Lorentz transformations, it nevertheless exhibits a symmetry which becomes 
 the standard Lorentz symmetry on shell.
The advantage of the ``doubled'' formalism 
is that  details of the duality group are not important for its quantum realization 
 -- the main features  are the same 
for discrete or continuous 
 duality symmetries. 
Since the ``doubled'' action is manifestly duality invariant it should  be possible to maintain it 
in the presence of a UV regularization. 
If the regularization also preserves the off-shell Lorentz-type  symmetry
present in the  classical action  then the on-shell
observables, such as S-matrix elements, should  exhibit both the duality and the 
Lorentz invariance.

While the  discussion in the previous sections focused  mainly on 
 two-dimensional examples, it is straightforward  to extend it to 4
dimensions. 
For example, it is straightforward  to construct the ``doubled''
 action for the theory \rf{1} similar to the one  describing the scalar-vector sector 
of ${\cal N}>4$   supergravity in 4 dimensions. We may either follow the strategy described in 
sec.~\ref{ads_classical} and  start  with  the first-order phase-space 
 action or   construct a  first-order ``master action"
 (as discussed  e.g. 
in Appendix~A of \cite{rot1}) 
 whose gauge symmetry may then be fixed in a convenient way.  
Dropping total derivatives (and fixing $A_0=0$), the result is \ci{ss} ($i=1,2,3$)
\be
&&\hat S =  \int d^4 x\Big[ - \ha (\del_a \p)^2 -
 \ha e^{4 \p} (\del_a \cc)^2   +\hat  L(A,\tA; \p, \cc)  \Big] \ , \ \\
&&\hat  L=\ha \Big( \rE^T_i {\cL} \rB_i 
      - \rB^T_i M \rB_i \Big)  \ , 
 \la{4.1}
\ee 
where\foot{This  action is similar the 2d one in \rf{1.66}.
 Note that in  $d=2p+2$  dimensions 
${\cL}=
\begin{pmatrix}
 0 & 1 \cr
(-1)^p & 0
\end{pmatrix}$. }
\be 
&& \rE_i = \partial_0  \rA_i \ , \ \ 
\qquad  
\rB_i = \epsilon_{ijk} \partial_j \rA_k\ , \ 
\qquad  
\rA_i = \begin{pmatrix}  A_i \cr
\tA_i 
\end{pmatrix}   \ ,\la{eb}  \\  
&&
{\cL}=
\begin{pmatrix}
 0 & 1 \cr
-1 & 0
\end{pmatrix}     
\ , \ \ \ \ \ \ \ \ \ \ \ \ \ \ 
M=
\begin{pmatrix}
e^{-2\phi} +4\chi^2 e^{2\phi}& -2\chi e^{2\phi} \cr
-2\chi e^{2\phi}              & e^{2\phi}
\end{pmatrix}    \ .   \la{lm} 
\ee
In  the  case of  $\chi=0$, this action is invariant under the  $Z_2$ 
duality transformation 
($\cL^T=- \cL, \   \cL^2 =- I$) 
\be
 \rA'_i = \cL \rA_i\ , \  \ \ \  M'=  \cL^T M \cL \ , \ \ \ \ {\rm i.e.} \ \ \ \
A_i ' = {\tilde A}_i
~,\quad
{\tilde A}'_i =  -A_i 
~,\quad 
\phi'= -\phi
\ .
\label{d4d}
\ee
\def \ri {{\rm i}}
\def \rF {{\rm F}}
The corresponding equations of motion are 
$E_i - e^{2 \p} \td B_i=0, \ \ \td E_i  + e^{-2 \p}  B_i=0$.
The  action \rf{4.1}  has  also a modified Lorentz-type  symmetry \ci{ss}
which becomes  the standard Lorentz symmetry on the equations of motion;
as in the 2d examples, 
the S-matrix elements should exhibit this symmetry simultaneously with being invariant under \rf{d4d}. 

Given that  $\cL^2=-I$  it is natural to introduce the complex combinations 
\be {\rm A}^\pm_i  \equiv  A_i \pm \ri{\tilde A}_i \ , \ \ \ \ \ \ \ \ 
\bar \rA^+_i = \rA^-_i \  ,   \la{aaa}  \ee
which  transform under the duality \rf{d4d} as 
\be 
(\rA^\pm_i )' = \mp \ri  \rA^\pm_i   \ ,\ \ \  \ \ \ \ \ \ \p '=-\p \la{tra}\ . \ee
The  classical equations 
written in terms of  derivatives of $\rA_i^\pm$ take the form 
\be \la{eqa}
\rE^+   +   \ri (  \rB^+  \cosh 2 \p   - \rB^- \sinh 2 \p) =0  \ , \ \ \ \ \ \ \ \ 
\rE^-   -   \ri (  \rB^-  \cosh 2 \p   -  \rB^+ \sinh 2 \p) =0  \ .
\ee
For $\p=0$ they become the (anti)self-duality conditions:
$\rF^\pm _{mn} = \pm \ri    \epsilon_{mn}^{\ \ \ kl} \rF^\pm _{kl}$, where as before $k,l,m,n=0,1,2,3$.
$\rA_i^\pm$  will  thus describe  on shell  photons of definite helicity
 (see \ci{gp,di} and below). 
The Lagrangian $\hat L$ in  \rf{4.1} written in terms of $\rA^\pm_i $ (leading to \rf{eqa})
becomes\foot{Note that the natural role of  complex combinations  
and the different structure of this action as  compared to  \rf{1.11},\rf{1.12}
is related to  the fact that real solutions of self-duality equations 
or chiral forms  exist  in dimensions $d= 4q+2=2,6,10,...$.}
\be
\hat L  =\fo 
 \Big[ {\rm i}  ( \rE^+_i  \rB^-_i  - \rE^-_i  \rB^+_i )
      -  2 \cosh 2 \p\ \rB^+_i  \rB^-_i    -     \sinh 2 \p\
  ( \rB^+_i  \rB^+_i  +\rB^-_i  \rB^-_i )  \Big] \ , \la{4.2}
\ee 
and is obviously  invariant under \rf{tra}.

This symmetry implies that the S-matrix elements without external $\phi$ lines 
labeled by $\rA^\pm$ fields,  $({\rA}^+)^{n_+}({\rA}^-)^{n_-}$, 
 pick up a phase $i^{-n_+ + n_-}$
under the duality \rf{tra};    since they must be invariant,  they 
 are nonvanishing only if 
$n_+-n_-=4k$.


One may repeat the above discussion  for the 4d Born-Infeld  theory
(for simplicity we set  coupling to  scalars   $\p,\cc$ to zero)
\be 
L(A)= - \sqrt{ 1 + \ha  F^{mn}F_{mn}  - {\textstyle{1 \ov 16}} (F^{mn} F^*_{mn})^2}   \ , \la{bin}\ee
which is  semiclassically equivalent to the following quadratic   in  $F_{mn}$  Lagrangian \ci{rot1}
involving two real auxiliary fields $U,V$ \foot{As in the scalar case (cf.\rf{hh},\rf{ll}), one may prefer  to use 
the ``polynomial'' action \rf{uv}  in the path integral definition of the quantum BI  theory.}
\be 
\la{uv}
L(A; U,V)= - \four \Big(  V F^{mn}F_{mn}  -    U  F^{mn}F^*_{mn}\Big)   - \ha \Big(  V + V^{-1}  + V^{-1} U^2 \Big) 
\ . \ee
The ``doubled'' action for  \rf{uv} is found  in the same way as \rf{4.1}
and is quantum-equivalent to \rf{uv}.  
The ``doubled'' action  for the BI  theory \rf{bin}   found from its  phase-space  formulation
follows also from the ``doubled''  action  for \rf{uv}  upon 
eliminating $U,V$ through 
their equations of motion.  The resulting  non-polynomial   ``doubled''  Lagrangian corresponding to 
\rf{bin}   written in terms of  derivatives of $\rA^\pm_i= A_i + \ri \td A_i $ fields 
is \cite{rot1} 
\be
\hat L&=& 
\ha ( E_i \td B_i - \td E_i B_i) - \sqrt{ 1  +  B_i^2 + \td B_i^2  + B^2_i \td B^2_k  - (B_i \td B_i)^2} \no \\
&=& \fo  {\rm i}   (\rE^+_i \rB^-_i - \rE^-_i \rB^+_i)
       -\sqrt{1+\rB^+_i\rB^-_i +\fo  (\rB^+_i \rB^-_i)^2 -\fo  (\rB^+_i\rB^+_i)(\rB^-_k\rB^-_k)} 
 \ .
\label{4.9}
\ee
Expanding the square root,  one finds that  to quadratic order this 
is the same as the  ``doubled'' action \rf{4.2} (with $\phi=0$).
Let us note that as  in the non-linear scalar  theory case (cf.  \rf{kk} and \rf{pl})  the  Lagrangian \rf{4.9} 
 is a priori equivalent to \rf{uv} 
only semiclassically (i.e. at the tree and one-loop level)  since the integral over $\tA_i$  (or $\tilde B_i$) 
here is non-gaussian.

 The Lagrangian~\rf{4.9} is invariant 
under the {\it same}  duality  transformation \rf{d4d}.\foot{Note that 
the  form of the on-shell relation  between the dual and original field strengths is 
of course modified, so in this sense  one may say that  BI equations of motion are covariant with respect to ``deformed'' version of Maxwell theory duality. This distinction is absent in the ``doubled'' description.} 
The  consequences of this symmetry  for 
the scattering amplitudes are also  the same:
the difference between the number of positive and negative helicity photons, $n_+-n_-$, must be a multiple of 4 (in particular, zero).
In fact,  the    S-matrix corresponding to   \rf{4.9} 
is actually helicity-conserving (see \ci{ros,bo}
for earlier demonstrations of this  for the  BI theory). Indeed,
every term in the expansion of
\rf{4.9} has an equal number of $\rB^+$ and $\rB^-$ factors,  and since the propagator 
connects $\rB^+$ and 
$\rB^-$,  in every Feynman graph (at any loop  order) the numbers 
of external positive and negative helicity photons (i.e. the 
numbers of on-shell $\rB^+$ and $\rB^-$  fields) 
are equal.
The  fact that the 
structure of the amplitudes appears to be  more constrained than required by the 
duality \rf{4.2} is not too surprising:  since there exist
infinitely many generalizations of the Maxwell's action exhibiting
duality (and thus having  an S-matrix obeying the $n_+-n_-=4k$ rule), 
the additional constraints implying helicity conservation\foot{Note that helicity 
conservation is a feature of some supersymmetric theories \ci{gp,di}; 
 the BI action admits a natural supersymmetrization  \ci{cf}.}
 reflect   the special property of the Born-Infeld action.\foot{Apart from the discrete duality \rf{d4d},
  the action \rf{4.9}  exhibits also  the continuous symmetry
$(\rA^\pm_i)'=e^{\pm {\rm i}\alpha} \rA^\pm_i$, with \rf{d4d} being included as the special case of 
$\alpha=-{\pi\ov 2}$. One may then understand \cite{ros} the additional 
restriction $n_+-n_-=4k=0$ as a consequence of the  infinitesimal part  of this symmetry. This infinitesimal 
symmetry is not present for the action in \rf{4.2}. 
Let us mention also  that the action of linearized $SL(2,R)$ invariance
on S-matrix of scalars and vectors for D3-brane 
in tree-level open/closed string theory 
 was  studied  in \ci{gar}.}

It is interesting to contrast the emergence of helicity conservation in the ``doubled''   theory  \rf{4.9} and in the standard  (Lorentz-covariant) description  
of  the Born-Infeld theory  in which the action \rf{bin}  contains interacting 
terms of the type $(F^+{}^2)^m (F^-{}^2)^n$ with 
arbitrary $m\geq 1 $ and $n\geq 1 $. Here $F^\pm$ are the self-dual and anti-self-dual components of the field strength which,  
similarly to $\rB^\pm$, reduce on shell to the positive/negative helicity photons.  At the  tree-level,
the contribution of the $m\ne n$ terms to helicity non-conserving amplitudes is cancelled \cite{bo} by 
Feynman graphs with contact (momentum-independent) propagators 
$\langle F^+(p)F^+(-p)\rangle$ and $\langle F^-(p)F^-(-p)\rangle$. 
In the ``doubled''  theory \rf{4.9}  all these cancellations are built into  the action.
The remaining graphs, in which all propagators are
 $\langle F^+(p)F^-(-p)\rangle\sim {1\ov p^2}$, generate 
only helicity-conserving amplitudes. 


Finally, let us comment on some implications for extended supergravities.
The ${\cal N}=8$ supergravity may be obtained by compactifying type
IIB 10d supergravity on a 6-torus;
the corresponding $O(6,6)$ symmetry is a subgroup of the full
$E_{7(7)}$ duality group. The physical states
of the theory are in one-to-one correspondence \cite{GM2} with the
states of the doubleton multiplet of
the maximally extended superconformal group in 4 dimensions,
$SU(2,2|8)$. While
interactions break
this symmetry to the maximally-extended Poincar\'e superalgebra, the
physical fields
continue to transform in representations of $SU(8)$
(it is possible to argue that all four-dimensional generalized unitarity cuts of all S-matrix
elements to any loop order
are $SU(8)$-invariant; one may therefore expect that the S-matrix has
this symmetry).
It turns out\footnote{We thank Murat Gunaydin for discussions on this
point.} that
the complete duality group is the closure of the six commuting $Z_2$
subgroups of $O(6,6)$ together with the
$SL(2,R)$ symmetry of the type IIB supergravity and this  $SU(8)$ symmetry.
While our  discussion in this paper  focussed  on  a  simple  example
of discrete
$Z_2$ duality  similar   considerations   should  apply also to the
full duality
 symmetry of the  ${\cal N}=8$ supergravity.\foot{Let us  note   that
as was argued in \ci{marcus} and more explicitly in \ci{bos} the
 $SU(8)$  chiral anomalies that would imply \ci{bos}  a breakdown of
$E_{7(7)}$ duality in the quantum theory  (the $E_{7(7)}$ anomaly is
determined by the $SU(8)$ anomaly through a Wess-Zumino consistency
condition \ci{bos}) actually cancel out.
Our focus here was not on the possible anomaly aspect but rather on the
realization of the duality in the quantum theory.
Let us note also  that
while the global $SU(8)$ is only an on-shell symmetry, it is possible
to formulate ${\cal N}=8$
supergravity in such a way  that the Lagrangian has $SL(8,R)$
\cite{jul0} or $SU^*(8)$ \cite{Hull:2002cv}
off-shell symmetry.
The $E_{7(7)}$ duality group is the closure of either one of these
groups together with  the 10-dimensional $SL(2,R)$
symmetry of type IIB supergravity and six abelian generators (2
non-compact and 4 compact).
Considerations similar to ours should apply, e.g.,
 to the  $SL(2,R)$  symmetry as well as  to the two non-compact
of the extra six abelian generators  (these are  also generators  of $O(6,6)$); see
\cite{Gunaydin:2009zza,
Gunaydin:2000xr} for the relevant change of basis for $E_{7(7)}$ generators.}

Our discussion in this paper   suggests that   the S-matrix
and the associated on-shell counterterms
computed in perturbative loop expansion\foot{Note that to relate the leading divergence of the 
effective action $\G= \sum_n \G_n$ (first  appearing at some $n$-th loop order) to 
to the corresponding divergent term in the S-matrix one  needs to evaluate $\G_n$ 
on a ``scattering''  solution of  just  classical (un-corrected) equations.  For the 
same reason, being interested only in the duality properties of the leading 
counterterms,  one does not need   to worry about  modification of the duality
transformation by finite quantum corrections.}
%
%
should be invariant under the $E_{7(7)}$ transformations of the scalar fields
 together with duality
transformations acting on the chiral (anti/self-dual) parts of the
vector fields. The latter symmetries are manifest  in the ``doubled"
formulation of the theory, 
where the action is not  invariant under the standard (tangent-space)
Lorentz symmetry; nevertheless, as we have argued in examples above,
 the  on-shell  effective action and/or the
 S-matrix  should have this  symmetry  along with the  duality symmetry.
It remains an open question
 whether there  may be  some additional implications of the  $E_{7(7)}$
 duality for the structure of potential  counterterms, as
conjectured in \ci{rk}.

\section*{Acknowledgments }

We would like to thank  R. Kallosh   for  illuminating 
discussions and explanations. RR  is  grateful to M. Gunaydin  for useful 
discussions  on  symmetries of   ${\cal N}=8$  supergravity. 
AAT is grateful to J. Buchbinder and N. Pletnev 
for a   discussion related to  4d case in Appendix A. 
The  work of RR  was supported by the National Science Foundation under grant PHY-08-55356.
The work of AAT was supported by the ERC Advanced  grant No.290456.


 \

\appendix

\section*{Appendix A:  Issue of quantum  
 $\p\to -\p$ invariance \\  \ \ \ \ \ \ \ \ \ \ \ 
\ \ \ \ \ \   on  curved background   }

\refstepcounter{section}
\def\theequation{A.\arabic{equation}}
\setcounter{equation}{0}

Integrating $x,\tx$ out in \rf{1.3},\rf{1.5} one expects to find the $\p \to -\p$ symmetry 
in the resulting effective action. This is not, however, automatic  if other fields  and symmetries 
are present and may depend on a regularization prescription (reflected in a choice of 
finite local counterterms).
To illustrate this, let us consider the following  example with  just one 2d scalar   field $x$
coupled to an external 2d scalar $\p$ and  2d metric $g_{ab}$ \foot{In  this Appendix we assume the world-sheet signature to be euclidean.}
\be 
\Gamma[\phi,g_{ab} ]
 =- \ln \int [dx]\  e^{- {1 \ov 2} \int d^2 \s \sqrt g   g^{ab}  G  \del_a
 x \del_b x } \ , 
\ \ \ \ \ \  \ \ \ \ \ \ G\equiv  e^{-2\p} \ .
\la{gt}
\ee
In the string or 2d sigma model context  one 
may think of $G$ as  a component of a target space metric in isometric 
direction $x$ (cf. \rf{1.2}).
 2d on-shell duality implies $G\to G^{-1}, \ x \to \td x$, with 
$ G \sqrt g   g^{ab}  \del_a x = i \ep^{ab} \del_b  \td x$. 
As was shown in \ci{st},  the definition of $\G$ in \rf{gt}  implies 
\be 
\Gamma[\p,g_{ab} ] -  \Gamma[-\p,g_{ab} ] = {1 \ov 8 \pi} \int d^2 x \sqrt g \ \p  R   \ ,  
\la{2d}
\ee
where $R$ is the curvature of $g_{ab}$.
This means, in particular, that under T-duality $G \to G^{-1}$   the target-space dilaton 
get shifted    \ci{bu}  by $\p= - \ha \ln G$ term. 
In  the  present context we may interpret \rf{2d} as an anomaly (present only in a curved 2d background)  of the
$\p \to - \p$   symmetry.
 More precisely, since \rf{2d}
 is  a local term, one may interpret it  not as a genuine  anomaly 
but as a 
finite local  counterterm required  for preservation of some  other 
symmetry  -- target space reparametrization  invariance in this 2d sigma model context. 
 As this counterterm breaks $\p \to -\p$ symmetry,
 that means that both symmetries --
the  2d duality  and the target space reparametrization  invariance -- cannot be manifest 
at the same time.\foot{To recall, anomalous, i.e.  symmetry 
   violating terms are usually (i)  nonlocal
and   (ii)   depend on which symmetry
of two or more one wants to preserve, i.e.   depend on a quantization prescription.
   In some cases  anomalous terms may be local, but then they are ambiguous,
    i.e. can be altered by adding 
finite local counterterms. For example,  in 4d conformal anomaly case the stress-energy tensor trace 
$T^m_m $  may contain a  total derivative term $ D^2 R$   that
  corresponds to  non-Weyl-invariant (finite)  local term   $R^2$ in the 
effective action. As such term can be cancelled by a local counterterm, 
the   $D^2 R$ term in the stress tensor anomaly is
ambiguous (may  depend on a  gauge choice  in the vector field  case,  etc).
In 2  dimensions there are similar terms that are cancelled   by
introducing a local  $R$-counterterm, i.e. the 
dilaton coupling  $ \int d^2 \s  \sqrt  g  R \Phi(x)$  which    cancels  the derivative
terms  in $T^m_m$.
In general,  one is required to introduce all possible local counterterms
and try to satisfy Ward identities of   required   symmetries; 
in the anomalous case this can be done only for a subset of all classical symmetries.}

Indeed, let us recall  the assumptions  that went into the derivation of \rf{2d}  in \ci{bu,t,st}.
It was assumed  that the path integral  measures used  for  all of 
 the fields -- the  original scalar, the dual scalar  and the 
2d auxiliary vector $n_a$ (needed to perform the  duality transformation at  the 
path integral level) 
contain the  same factors of $G$,  i.e.  are covariant 
with respect to the  target space metric. This follows from the requirement of 
target space reparametrization invariance which is natural  in the sigma model context.
 As the right-hand side  of  \rf{2d} 
 is  a $local$ term,  it depends effectively on  a choice of a quantization scheme 
or regularization prescription. Indeed, one  way to get this expression 
 is to notice  \ci{bu}
that the path integral over $n_a$ with action $\int d^2 x \sqrt g \  G  n^a n_a$ 
gives the following contribution to $\G$
\be  \ha \tr \ln ( G e^{- \Lambda^{-2}  \Delta_1} ) = \ha \int d^2 x \sqrt g \  \ln G\ 
  \Big[ 2\Big(  \Lambda^2 + {1 \ov 6} R\Big)  - R \Big] \ , \ee
where $(\Delta_1)_{ab}  =   - g_{ab} \nabla^2 + R_{ab}  $  is a natural operator on vectors used to  regularize the local ``$\delta(0)$'' factor  and $\Lambda $ is a UV cutoff.
 $2 ( \Lambda^2 + {1 \ov 6} R) $ term  cancels against similar terms from the 
 scalar and dual scalar contributions to $\G$  but  $-R$ term survives  and leads to \rf{2d}.

Let us  now consider the corresponding 4d example \rf{1}
on a curved 4d background, i.e. define 
\be 
\Gamma[\phi,g_{mn} ] =- \ln \int [dA]\  e^{- {1 \ov 4} \int d^4 x \sqrt g g^{mn} g^{pq}\  e^{-2\phi} F_{mp}   F_{nq}}  \ . 
\la{ga}
\ee
Here 
the  classical equations  of motion 
 have symmetry 
under $A\to \td A, \ \p \to - \p$  with $e^{-2\phi} *dA = d \td A$
and one may ask if this 
classical symmetry becomes  symmetry of the  effective action \rf{ga}, i.e.
$ 
\Gamma[\phi,g_{mn} ] = \Gamma[-\phi,g_{mn} ]
$.
This is, of course, expected from formal path integral transformation argument implying 
 that $\G$ should depend only on derivatives 
of $\p$ and only on  even powers of $\p$.\foot{$SL(2,R)$ invariance of the 
conformal anomaly (controlling  logarithmically 
 UV divergent  part of the effective action 
and thus its finite Weyl-anomalous part)
resulting from  integrating over the vector  field  coupled to $(\p, \chi)$ 
as in \rf{1} was explicitly demonstrated in \ci{osb}.}
 For example, if  $\G$  were to 
contain a local  term 
$
\G' =  a_0 \int d^4 x \sqrt g \  \phi\   C^2_{mnkl} \ , 
$
where $C_{mnkl}$ is the Weyl tensor then 
this term would   change under the constant shift of $\phi$ but such shift in  \rf{ga},
but in  view of the above 2d  example one  may  avoid this objection by replacing $C^2_{mnkl}$
by the 4d Euler density combination.

Indeed, this is what we  find if we follow the same steps  that in 
2d case led to \rf{2d}.
In  general, torison of an elliptic complex\foot{This is  a set of operators  like the three 
 ones  -- scalar, vector, and dual scalar -- mentioned in 2d case.} will be given by a
combination of  Seeley coefficients (the one appearing in  $t^0$  power  in expansion of $
\Tr {\rm e}^{-t \Delta} = \sum_k B_k t^k$). It is  straightforward 
 to repeat the analysis in \ci{st}
for the torsion $\ha \sum^d_{n=0} (-1)^n (n+1) \ln \det \Delta_n$  of the 
4d elliptic complex (scalar, vector and 2-tensor operators). This will give the  
 analog of \rf{2d} originating 
 from the corresponding Seeley coefficients $b_4 \sim R_{mnkl}^2 + ...$. 
 Under the same $assumption$ as in the 2d case  that  all the  measure factors are the 
 same for all the operators in the complex,  
 this computation was done in  \ci{gil} and  led to the 
direct analog of \rf{2d}  with the  4d Euler density replacing the 2d one: 
\be 
\Gamma[\p,g_{mn} ] -  \Gamma[-\p,g_{mn} ] = -{1 \ov 32 \pi^2} \int d^2 x \sqrt g \  \p  
(R^2_{mnkl} - 4 R^2_{mn} + R^2) \ . 
\la{4d}
\ee
Given that  the Euler density is a total derivative, this expression depends only on
 $\del \p$ (assuming 
trivial  topology). 
As this is a local  term, one may interpret  its presence as 
reflecting   the desire to preserve some other symmetry at the expense 
of the duality $\p \to -\p$. 
  Since, in  contrast to 2d sigma model case,   in the 
  4d vector case of \rf{1} we do not have
target space diffeomorphisms 
 acting on vectors
  we may instead insist  on the preservation of 
 the duality by  fine-tunning the coefficient of this term to zero, i.e. by canceling it   by 
  a  local counterterm.\foot{This 
   may 
   be required in the context of coupling this model to gravity, but 
in the supergravity context one may expect measure-related  factors to cancel.
This may, of course, also depend on a choice of   field redefinitions
relating classically equivalent    supergravity theories. 
In fact,   similar  $\int d^2 x \sqrt g \  \p  R^2_{mnkl} $
local counterterm previously appeared in \ci{gri} 
when discussing the quantum  equivalence of the $SO(4)$ and $SU(4)$  versions of ${\cal N}=4$
supergravity. It is directly related to the fact that the required 
field redefinition \ci{jul} produces the duality-anomalous Jacobian.
}

Let us  also note 
that coupling an ${\cal N}=4$  vector  supermultiplet  (SYM) to ${\cal N}=4$ conformal 
supergravity \ci{deroo}  and integrating out  the ${\cal N}=4$   vector  multiplet fields 
leads to an effective  action \ci{lt}  whose UV divergent  part  should be 
the same as the action of ${\cal N}=4$ conformal supergravity \ci{ber,ft}
which  should have  manifest 
 {\it off-shell}   $SL(2)$ duality  invariance involving the dilaton $\p$ and  its
  pseudoscalar partner $\cc$, 
with  $SU(4)$ vectors  $not$  transforming (here there is no 
scalar-vector coupling, in contrast to the case of 
  ${\cal N}=4$ Poincar\'e supergravity). 
  The invariance of the full   local part of the resulting quantum effective action  under $\p\to - \p$ 
will thus hold provided  the  ``anomalous'' term in \rf{4d} is  cancelled  by a
 local counterterm.
 
 As was noted in the introduction, 
 in addition to the above   local   non-invariant terms that can be    removed   by  local counterterms, 
 the effective action of a theory like \rf{1}   contains also genuine non-local duality non-invariant (anomalous)  
 terms   containing scalars   and  curvature-dependent  $RR^*$  factor (cf. \ci{dolg}).    They cancel 
 only in ${\cal N}=8$   supergravity \ci{marcus}.


\section*{Appendix B: One-loop effective action corresponding to\\ scalar theory \rf{ll}}

\refstepcounter{section}
\def\theequation{B.\arabic{equation}}
\setcounter{equation}{0}

The quadratic operator in eq.~(\ref{kkk}) has constant coefficients ($n_a= \del_a x$=const, 
\ $ G_\xx=(1 + n^2)^{-1/2} $=const)  so that we find 
($V_2$ is the 2d  space-time volume factor)
\be
\G_1 &=& \ha V_2 \int \frac{d^2q}{(2\pi)^2} \ln\Big(G^{-1}_\xx q^2 -G_\xx (n\cdot q)^2 \Big)\cr
&=&\ha V_2 \int \frac{d^2q}{(2\pi)^2} \Big(- \ln G_\xx+\ln q^2
 +\ln\Big[1 - G_\xx^2 \frac{(n\cdot q)^2}{q^2}\Big]\Big) \ . 
\label{B2}
\ee
The 
 second term here  may be dropped as it is independent of the classical background.
 Explicitly, 
 \be
\ha \int \frac{d^2q}{(2\pi)^2} \ln\Big[1 - G_\xx^2 \frac{(n\cdot q)^2}{q^2}\Big]=
-\ha \sum_{k=1}^\infty\frac{1}{k}\frac{(2k-1)!!}{(2k)!!}(G_\xx^2\,n^2)^k\; I = 
\ln[ \ha  (1+G_\xx)]\;I \ , 
\label{B4}
\ee
 where we used that 
\be
\int\frac{d^2q}{(2\pi)^2} \left(\frac{(n\cdot q)^2}{q^2}\right)^k=\frac{(2k-1)!!}{(2k)!!}(n\cdot
 n)^k\;I \ , \ \ \ \ \ \ \ 
 I\equiv \int {d^2q\ov (2\pi)^2}  =\Lambda^2 \ , \la{B3}
 \ee
 and that $ G_0 = (1 + n^2)^{-1/2}$.
Then the resulting  combination in the integrand of \rf{B2} is 
 \be  - \ha \ln G_\xx +  \ln[ \ha  (1+G_\xx)]  
 =  \ln[ \ha  (G_\xx^{1/2} +G_\xx^{-1/2})]    \ ,
 \la{jj}
 \ee  so that 
 the function $F$ in \rf{2.21} is 
\be 
F= \ln[ \ha  (y^{1/2} +y^{-1/2})] \ , \ \ \ \ \ \ \ \ F(y)= F(y^{-1})
\ .\la{hhh} \ee

\newpage

\bigskip

\end{document}